\documentclass[preprint, 3p]{elsarticle}

\usepackage{amsmath,graphicx}

\usepackage{amsmath,graphicx}
\usepackage{amssymb,amsmath,epsfig,url,mathrsfs} 
\usepackage{algorithm,algorithmic}
\usepackage[usenames, dvipsnames]{color}
\usepackage[framemethod=TikZ]{mdframed}
\usepackage{xcolor}
\usepackage{listings}
\usepackage{subcaption}

\usepackage[norelsize,algo2e]{algorithm2e} 
\SetKwInput{KwInput}{Input} 

\usepackage{scalerel}

\usepackage{arydshln}
\usepackage{tikz}
\usepackage{amsfonts}
\usepackage{amsmath,amssymb}
\usepackage{systeme,mathtools}
\usetikzlibrary{positioning,arrows.meta,quotes}
\usetikzlibrary{shapes,snakes}
\usetikzlibrary{bayesnet}
\tikzset{>=latex}
\tikzstyle{plate caption} = [caption, node distance=0, inner sep=0pt, below left=0pt and 0pt of #1.south east]

\usepackage{ragged2e}
\usepackage{array}
\usepackage{booktabs}
\usepackage{siunitx}
\newcolumntype{L}[1]{>{\raggedright\let\newline\\\arraybackslash\hspace{0pt}}m{#1}}
\newcolumntype{C}[1]{>{\centering\let\newline\\\arraybackslash\hspace{0pt}}m{#1}}
\newcolumntype{R}[1]{>{\raggedleft\let\newline\\\arraybackslash\hspace{0pt}}m{#1}}

\mdfdefinestyle{MyFrame}{%
    linecolor=black,
    outerlinewidth=.5pt,
    roundcorner=5pt,
    innertopmargin=\baselineskip,
    innerbottommargin=\baselineskip,
    innerrightmargin=5pt,
    innerleftmargin=0pt,
    backgroundcolor=gray!10!white}
\definecolor{codegreen}{rgb}{0,0.6,0}
\definecolor{codegray}{rgb}{0.5,0.5,0.5}
\definecolor{codepurple}{rgb}{0.58,0,0.82}
\definecolor{backcolour}{rgb}{0.95,0.95,0.92}

\lstdefinestyle{mystyle}{
    backgroundcolor=\color{backcolour},
    commentstyle=\color{codegreen},
    keywordstyle=\color{blue},
    numberstyle=\tiny\color{codegray},
    stringstyle=\color{codepurple},
    basicstyle=\footnotesize,
    breakatwhitespace=false,
    breaklines=true,
    captionpos=b,
    keepspaces=true,
    numbers=none,
    numbersep=5pt,
    showspaces=false,
    showstringspaces=false,
    showtabs=false,
    tabsize=2
}

\usepackage{amsmath}
\usepackage{graphicx}
\usepackage[colorinlistoftodos]{todonotes}
\usepackage[colorlinks=true, allcolors=blue]{hyperref}

\graphicspath{{figures/}}

\DeclareMathAlphabet{\mathpzc}{OT1}{pzc}{m}{it}

\usepackage{multirow}
\usepackage{adjustbox, bm}
\usepackage{pifont}
\newcommand{\cmark}{\ding{51}}%
\newcommand{\xmark}{\ding{55}}%

\journal{Computer Speech \& Language}

\begin{document}

\begin{frontmatter}

\title{Deep Learning based Multi-Source Localization with Source Splitting and its Effectiveness in Multi-Talker Speech Recognition}

\author[jhu]{Aswin Shanmugam Subramanian\corref{cor1}}
\ead{aswin@jhu.edu}

\author[tai_s]{Chao Weng}
\ead{cweng@tencent.com }

\author[lti,jhu]{Shinji Watanabe}
\ead{shinjiw@ieee.org}

\author[tai_b]{Meng Yu}
\ead{raymondmyu@tencent.com }

\author[tai_b]{Dong Yu}
\ead{dyu@tencent.com }

\address[jhu]{Center for Language and Speech Processing, Johns Hopkins University, Baltimore, MD, USA}
\address[lti]{Language Technologies Institute, Carnegie Mellon University, Pittsburgh, PA, USA}
\address[tai_s]{Tencent AI Lab, Shenzhen, China}
\address[tai_b]{Tencent AI Lab, Bellevue, WA, USA}

\cortext[cor1]{Corresponding Author. A part of the work was carried out by this author during an intership at Tencent AI Lab, Bellevue, USA.}

\begin{abstract}
Multi-source localization is an important and challenging technique for multi-talker conversation analysis. This paper proposes a novel supervised learning method using deep neural networks to estimate the direction of arrival (DOA) of all the speakers simultaneously from the audio mixture. At the heart of the proposal is a source splitting mechanism that creates source-specific intermediate representations inside the network. This allows our model to give source-specific posteriors as the output unlike the traditional multi-label classification approach. Existing deep learning methods perform a frame level prediction, whereas our approach performs an utterance level prediction by incorporating temporal selection and averaging inside the network to avoid post-processing.  We also experiment with various loss functions and show that a variant of earth mover distance (EMD) is very effective in classifying DOA at a very high resolution by modeling inter-class relationships. In addition to using the prediction error as a metric for evaluating our localization model, we also establish its potency as a frontend with automatic speech recognition (ASR) as the downstream task. We convert the estimated DOAs into a feature suitable for ASR and pass it as an additional input feature to a strong multi-channel and multi-talker speech recognition baseline. This added input feature drastically improves the ASR performance and gives a word error rate (WER) of 6.3\% on the evaluation data of our simulated noisy two speaker mixtures, while the baseline which doesn't use explicit localization input has a WER of 11.5\%. We also perform ASR evaluation on real recordings with the overlapped set of the MC-WSJ-AV corpus in addition to simulated mixtures. 
\end{abstract}


\end{frontmatter}


\section{Introduction}
Direction of arrival (DOA) estimation is the task of estimating the direction of the sound sources with respect to the microphone array and it is an important aspect of source localization. This localization knowledge can aid many downstream applications. For example, they are used in robot audition systems \cite{hark, nakadai2001real} to facilitate interaction with humans. Source localization is pivotal in making the human-robot communication more natural by rotating the robot's head. The localization component in the robot audition pipeline also aids in source separation and recognition. Moreover, there is growing interest in using far-field systems to process multi-talker conversations like meeting transcription \cite{yoshioka2019advances}. Incorporating the source localization functionality can enrich such systems by monitoring the location of speakers and also potentially help improve the performance of the downstream automatic speech recognition (ASR) task. DOA estimation can also be used with smart speaker scenarios \cite{chime5is,haeb2019speech} and potentially aid with better audio-visual fusion. 

Some of the earlier DOA estimation techniques are based on narrowband subspace methods ~\cite{music_schmidt, espirit}.
The simple wideband approximation can be achieved by using the incoherent signal subspace method (ISSM), which uses techniques such as multiple signal classification (MUSIC) independently on narrowband signals of different frequencies and average their results to obtain the final DOA estimate. Broadband methods which better utilize the correlation between different frequencies such as weighted average of signal subspaces (WAVES) \cite{waves} and test of
orthogonality of projected subspaces (TOPS) \cite{tops} have shown promising improvements.  Alternative cross-correlation based methods like steered response power with
phase transform (SRP-PHAT) \cite{srp_phat} have also been proposed. Most subspace methods are not robust to reverberations as they were developed under a free-field propagation model \cite{chakrabarty2019multi}. Recently, there has been a shift in interest to supervised deep learning methods to make the estimation robust to challenging acoustic conditions.

The initial deep learning methods were developed for single source scenarios \cite{hirvonen_doa, takeda_doa, vesperini_doa, nelson_localization, chakrabarty2017broadband}. In \cite{takeda_doa} and \cite{vesperini_doa}, features are extracted from MUSIC and GCC-PHAT respectively. The learning is made more robust in \cite{hirvonen_doa, nelson_localization, chakrabarty2017broadband} by encapsulating the feature extraction inside the neural network. While \cite{vesperini_doa} treats DOA estimation as a regression problem, all the other methods formulate it as a classification problem by discretizing the possible DOA angles. The \textit{modeling resolution} which is determined by the output dimension of the network is quite low in \cite{takeda_doa}, \cite{hirvonen_doa}, and \cite{chakrabarty2017broadband} at $5^{\circ}$, $45^{\circ}$, and $5^{\circ}$ respectively. Although the modeling resolution is high in \cite{nelson_localization} with classes separated by just $1^{\circ}$, the evaluation was performed only with a block size of $5^{\circ}$ and it is also shown to be not robust to noise. In \cite{reg_vs_class}, it was shown that a classification based system with a grid resolution of $10^{\circ}$ and a regression based system gives similar performance.

Deep learning based techniques have been shown to be very effective for joint sound
event localization and detection (SELD) of overlapping sources consisting of both speech and non-speech events \cite{Adavanne2018_JSTSP, Shimada2021}. In \cite{Adavanne2018_JSTSP}, SELD is solved using a multi-task learning approach with DOA estimation treated as a multi-output regression while in \cite{Shimada2021}, detection and localization is solved jointly using a unified regression based loss. In this work, we focus only on localization of speech sources as we are interested in investigating the importance of DOA estimation as a frontend with speech recognition as the downstream task. 

Time-frequency (TF) masks estimated from speech enhancement \cite{irm_se} or separation \cite{dl_wang_target} systems have been used for DOA estimation to handle noise, and interference in \cite{zhang19j_interspeech, mack_icassp, zqwang_localize, ziteng}. The preprocessing speech enhancement system in \cite{zhang19j_interspeech, mack_icassp} is used to select time-frequency regions in the single speech source input signal that exclude noise and other non-speech interference. In \cite{ziteng} and \cite{sivasankaran}, the interference consists of a speech source so the preprocessing system takes an auxiliary input to extract the speaker of interest. In \cite{ziteng}, a target speech separation system similar to \cite{speaker_beam} is used and this preprocessing system takes pre-enrollment speech corresponding to the speaker of interest as an auxiliary input. In \cite{sivasankaran}, keywords were used to identify the target. Although, multiple source inputs are used in these methods, an independently trained preprocessing system is used to identify the target input and the source localization system performs a biased estimation based on this identifier. 

The speech enhancement systems used in single speech source scenario and target separation systems used in multi-talker scenario are typically trained with the groundtruth signals as the target and hence can be trained with only simulated data. Cascading with such a preprocessing system also increases the complexity of the overall DOA estimation framework. In \cite{zhang19j_interspeech}, a pseudo multi task learning based end-to-end training approach was proposed for a single speech source input. In end-to-end training both the speech enhancement network and the DOA estimation network were jointly optimized with only the DOA label (and without using the target clean signal). It was also shown to give comparable performance to a typical multi-task learning approach. End-to-end methods are interesting as it has the scope for training with real data as the DOA labels can be obtained with infrared sensors as used in \cite{locata}.

The deep learning models were extended to handle simultaneous estimation of multiple sources in an end-to-end manner by treating it as a multi-label classification problem in \cite{adavanne_ms_doa, chakrabarty2019multi, he_ms_doa}. These approaches perform an unbiased estimation to obtain the DOAs of all the sources. The multi-label classification approach still gives only a single vector as output with each dimension treated as a separate classification problem. This procedure is not good at capturing the intricate inter-class relationship in our problem. So increasing the number of classes to perform classification at a very high resolution will fail. During inference, the output vector is treated like a spatial spectrum and DOAs are estimated based on its peaks. In \cite{he_ms_doa, adavanne_ms_doa}, the number of peaks is determined based on a threshold. In \cite{chakrabarty2019multi}, the number of sources are assumed to be known based on which the peaks are chosen at the output. In \cite{adavanne_ms_doa} and \cite {chakrabarty2019multi}, the source locations are restricted to be only in a spatial grid with a step size of $10^{\circ}$ and $15^{\circ}$ respectively. To make the models work on realistic conditions, it is crucial to handle all possible source locations.  Moreover, these models perform a frame level prediction in spite of the sources assumed to be stationary inside an utterance. Hence, post-processing is required to get the utterance level DOA estimate.

In this work, we propose a \textit {source splitting} mechanism, which involves treating the neural network as two distinct and well defined components. The first component disentangles the input mixture by implicitly splitting them as source-specific hidden representations to enable tackling of multi-source DOA estimation via single-source DOA estimation. The second component then maps these features to as many DOA posteriors as the number of sources in the mixture. This makes the number of classification problems to be only equal to the number of sources and not the number of DOA classes like the existing methods. With the added help of loss functions that can handle the inter-class relationship better, our methods can classify DOA reliably at a high resolution by performing a fine-grid estimation. Although, source splitting formulates the network as two components, it is optimized end-to-end with only DOA estimation objective. Like \cite{chakrabarty2019multi}, we assume the knowledge of number of sources in the utterance but we incorporate it directly in the architecture of the model instead of using it while post-processing. 

 We evaluate our proposed localization method both as a standalone task and also based on its effectiveness as a frontend to aid speech recognition. Assuming the DOA to be known, the effectiveness of using features extracted from the ground-truth location for target speech recognition was shown in \cite{zhuo_loc, icassp-tencent} as a proof of concept. The importance of localization for multi-talker speech recognition was also shown in our previous work \cite{subramanian2020directional}. In \cite{subramanian2020directional}, we proposed a model named \textit{directional ASR}, which can learn to predict DOA without explicit supervision by joint optimization with speech recognition. In the second part of this paper, we devise an approach that can use localization knowledge to improve multi-talker speech recognition. We use a strong multi-talker speech recognition system called MIMO-Speech \cite{chang2019mimo, mimo_transformer} as a baseline. MIMO-Speech doesn't perform explicit localization of the sources. The DOAs estimated from our proposed models in the first part are converted to angle features \cite{zhuo_loc}. The angles features encode DOA information and act as a localization prior. These angles features are given as additional inputs to the MIMO-Speech model and tested for its effectiveness in improving speech recognition. It is hard to obtain DOA labels for real datasets. We  perform DOA estimation for real speech mixtures from multichannel wall street journal audio-visual (MC-WSJ-AV) corpus and evaluate the quality of localization solely based on the downstream ASR metric of word error rate (WER).

\section{Deep Learning based Multi-Source Localization}
\label{sec:doa}

\begin{figure}
\centering
\includegraphics[scale=0.35]{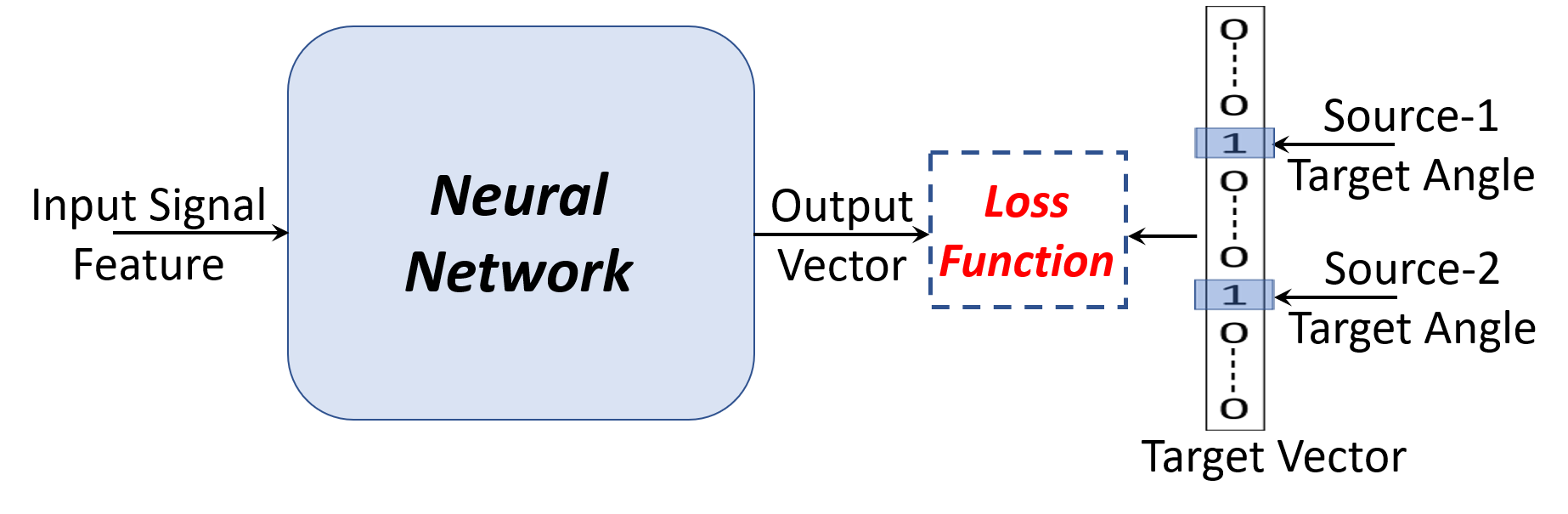}
\caption{Multi-label Classification Model for 2-source Scenario}
\label{fig:mlc_model}
\end{figure}

Given a multi-channel input signal with multiple sources, a deep neural network can be trained to predict all the azimuth angles corresponding to each of the sources. Typically it is treated as a multi-label classification problem \cite{chakrabarty2019multi, adavanne_ms_doa}, as shown in Figure \ref{fig:mlc_model}. Here, the network outputs only a single vector which is supposed to encode the DOA of all the sources. It is possible to have spurious peaks in angle classes that are in close proximity to any of the target angle classes, especially when the number of classes are increased. As more than one peak needs to be selected from the vector during inference, it is possible a spurious peak will also be chosen at the expense of a different source.  

\begin{figure}
\centering
\includegraphics[scale=0.35]{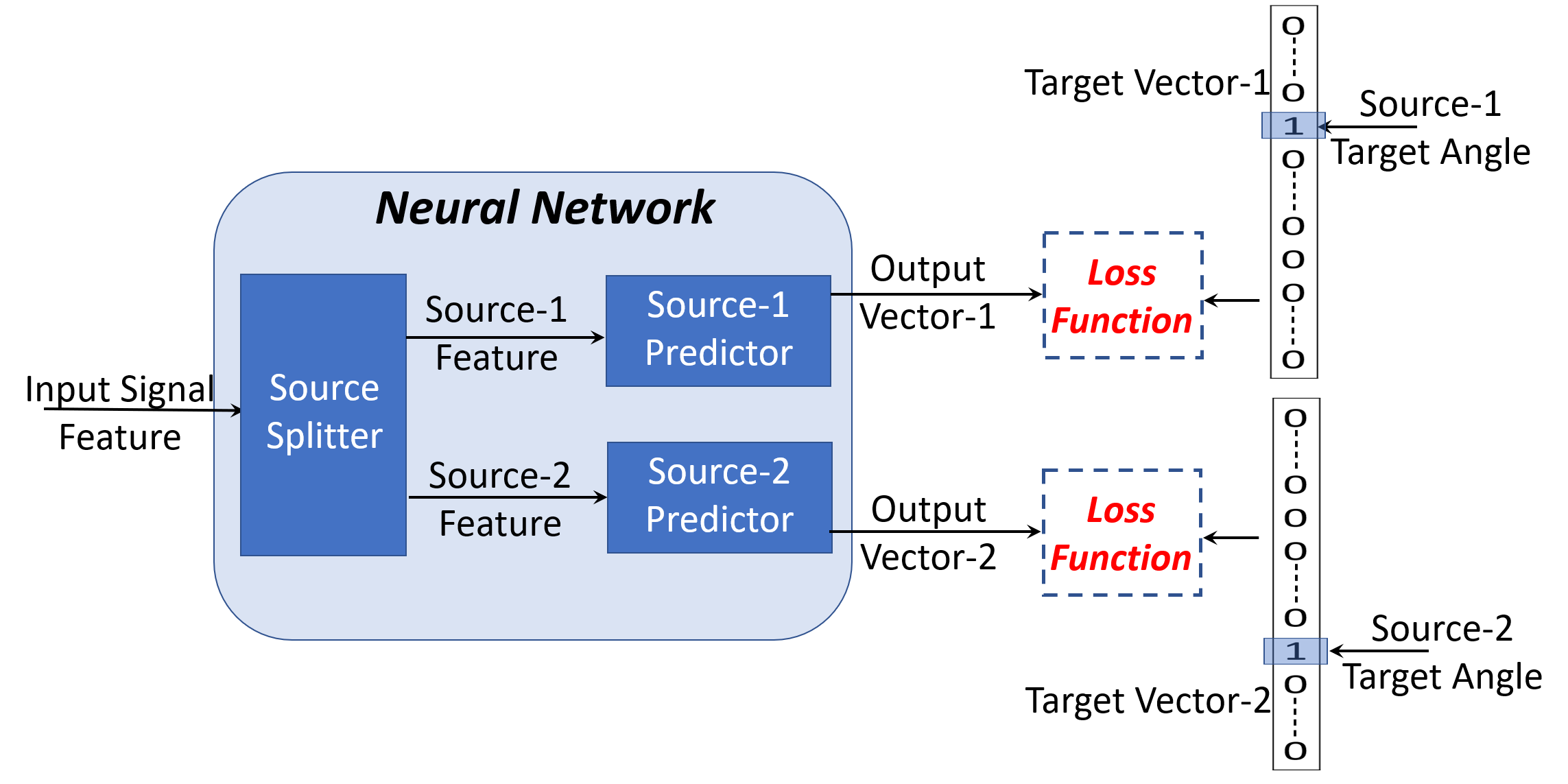}
\caption{Proposed Model with Source Splitting Mechanism for 2-source Scenario}
\label{fig:source_split_model}
\end{figure}

We propose an alternative modeling approach for multi-source localization. In this model, the network is divided into two well-defined components with specific functionalities as shown in Figure \ref{fig:source_split_model}.  The first component is the \textit{source splitter}, which dissects the input mixed signal to extract source specific spatial features. These features are passed to the second component named \textit{source dependent predictors}, which gives individual posterior vectors as output. As multiple outputs are obtained, each one has a well defined target with just one target angle corresponding to a specific source.

As we will have $N$ output predictions corresponding to the $N$ sources in our proposed model, there will be source permutation ambiguity in how we assign the predictions to the targets. We handle it in two ways. First we can use permutation invariant training (PIT) \cite{pit} that are popular for source separation models. In PIT, all possible prediction-target assignment permutations are considered a valid solution and the one with the minimum loss is chosen during training. Alternatively, we fix the target angles to be in ascending order and force only this permutation.

Although the sources are stationary in existing models like \cite{chakrabarty2019multi}, a frame level prediction is performed and then time averaging is used as a post processing step during inference. It is common in tasks like speaker identification, which involve converting a sequence of features into a vector to incorporate average pooling inside the network  \cite{xvector, seqsum}. We follow this approach in all our models to avoid post-processing.  

We describe the different deep learning architectures used in this work. First we describe an architecture based on multi-label classification. Then we introduce two different architectures based on our proposed source splitting mechanism. Let $Y_1, Y_2, \cdots, Y_M$ be the $M$-channel input signal in the short-time Fourier transform (STFT) domain, with $Y_m \in \mathbb{C} ^{T\times F}$, where $T$ is the number of frames and $F$ is the number of frequency components. The input signal is reverberated, consisting of $N$ speech sources. There can be additive stationary noise but we assume that there are no point source noises. We also assume that $N$ is known. In this work the array geometry is considered to be known at training time and only the azimuth angle is estimated.

\subsection{Multi-label Classification (MLC)}
\label{sec:cnn_blstm_mlc}
This model is designed to take the raw multi-channel phase as the feature. Let the phase spectrum of the multichannel input signal be represented as $\mathcal{P} \in [0, 2 \pi] ^{T\times M \times F}$. This raw phase $\mathcal{P}$ is passed through the first component of the localization network based on a convolutional neural network (CNN) given by $\text{LocNet-CNN}(\cdot)$ to extract phase feature $Z$ by pooling the channels as follows:

\begin{align}
     Z &= \text{LocNet-CNN} (\mathcal{P}) \in \mathbb{R} ^{T\times Q},
    \label{locnet_cnn}
\end{align}
where $Q$ is the feature dimension. The $\text{LocNet-CNN}(\cdot)$ architecture is inspired from \cite{chakrabarty2019multi}, which uses convolutional filters across the microphone-channel dimension to learn phase difference like features.

The phase feature $Z$ from Eq.~\eqref{locnet_cnn} is passed through a feedforward layer as follows,

\begin{align}
     W &= \text{ReLU}(\text{AffineLayer1}(Z)),
     \label{locnet_blstm_mlc}
\end{align}
where $W \in \mathbb{R}^{T \times Q}$ is an intermediate feature. We consider that the sources are stationary within an utterance. So in the next step a simple time average is performed to obtain a summary vector.

\begin{align}
    \label{summary_vector_mlc}
     \xi (q) &= \frac{1}{T} \sum_{t=1}^{T} w (t, q),
\end{align}
where $\xi (q)$ is the summary vector at dimension $q$, and $w(t, q)$ is the intermediate feature at time $t$ and feature dimension $q$.
The summary vector, represented in vector form as $\bm{\xi} \in \mathbb{R} ^{Q}$ is passed through a learnable $\text{AffineLayer}(\cdot)$ to convert its dimension to $\lfloor 360/\gamma \rfloor$, where $\gamma$ is the angle resolution in degrees to discretize the DOA angle. 

\begin{align}
     \label{mlc_vector}
     \bm{\kappa} &= \sigma(\text{AffineLayer2}(\bm{\xi})). \\
\end{align}
where, $\sigma(\cdot)$ is the sigmoid activation, and $\bm{\kappa} \in (0,1)^{\lfloor 360/\gamma \rfloor}$ is the multi-label classification vector. The DOAs can be estimated by finding the indices in $\bm{\kappa}$ corresponding to the $N$ largest peaks.

\subsection{Convolutional Mapping based Source Splitting Mechanism (Map-Split-C)}
\label{sec:cnn_blstm_map}
This architecture is a simple extension of the MLC model to achieve source splitting. Similar to the MLC model in Section \ref{sec:cnn_blstm_mlc} raw phase is used as the input and the phase feature $Z$ is extracted using Eq.~\eqref{locnet_cnn}. $Z$ will have DOA information about all the sources in the input signal. Source splitting is achieved through a source dependent affine transformation as follows,
 
\begin{align}
     W^{n} &= \text{ReLU}(\text{AffineLayer1}^{n} (Z)),
     \label{locnet_map}
\end{align}

Note that the only difference from Eq.~\eqref{locnet_cnn} is that the affine layer is made to be dependent on the source $n$ to achieve source splitting. Here, $W^{n}$ are intermediate representations specific to source $n$. These representations are averaged to create the source specific summary vectors in the following way,

\begin{align}
    \label{summary_vector_map}
     \xi^{n} (q) &= \frac{1}{T} \sum_{t=1}^{T} w^{n} (t, q),
\end{align}
where $\xi^{n} (q)$ is the summary vector for source $n$ at dimension $q$. 

The summary vector, represented in vector form as $\bm{\xi}^{n} \in \mathbb{R} ^{Q}$ is passed through a learnable feedforward layer $\text{AffineLayer2}(\cdot)$, which acts as the predictor and converts the summary vector from dimension $Q$ to the dimension $\lfloor 360/\gamma \rfloor$, where $\gamma$ is the angle resolution in degrees to discretize the DOA angle. Based on this discretization, we can predict the DOA angle as a classifier with the softmax operation. From this, we can get the source-specific posterior probability for the possible angle classes as follows,
     
\begin{align}
     \label{posterior_1}
     [\Pr (\theta^{n}=\alpha_{i}|\mathcal{P})]_{i=1}^{\lfloor 360/\gamma \rfloor} &= \text{Softmax}(\text{AffineLayer2}(\bm{\xi}^{n})), \\
     \label{class_angles_1}
     \alpha_{i} &=  \left( (\gamma * i) -\left((\gamma-1)/2\right) \right) \left(\pi/180\right),
\end{align}

Note that the parameters of the predictor are shared across the sources in Eq.~\eqref{posterior_1} but it is also possible to make this feedforward layer source dependent. The estimated DOA $\hat{\theta}^{n}$ for source $n$ is determined by finding the peak from the corresponding posterior in Eq.~\eqref{posterior_1} as follows,

\begin{align}
     \label{argmax_1}
     \hat{\theta}^{n} &= \operatorname*{argmax}_{\alpha_{i}} \Pr (\theta^{n}=\alpha_{i}|\mathcal{P}), 
\end{align}

\subsection{Masking based Source Splitting Mechanism (Mask-Split)}
\label{sec:cnn_blstm}
This architecture is inspired from the localization subnetwork used in directional ASR \cite{subramanian2020directional}. This model also uses raw phase as the input and the phase feature $Z$ is extracted in the same way as Section \ref{sec:cnn_blstm_mlc} using Eq.~\eqref{locnet_cnn}. $Z$ will have DOA information about all the sources in the input signal. It is processed by the next component $\text{LocNet-Mask}(\cdot)$ which consists of BLSTM layers. Source splitting is achieved through this component which extracts  source-specific ratio masks as follows,
 
\begin{align}
     [W^{n}]_{n=1}^{N} &= \sigma(\text{LocNet-Mask} (Z)),
     \label{locnet_mask}
\end{align}
where $W^{n} \in [0, 1]^{T \times Q}$ is the feature mask for source $n$ and $\sigma(\cdot)$ is the sigmoid activation. This mask segments $Z$ into regions that correspond to each source. Note that $\text{LocNet-Mask}(\cdot)$ can also implicitly perform voiced activity detection as it outputs ratio masks. 

 The extracted phase feature mask from Eq.~\eqref{locnet_mask} is used to perform a weighted averaging of the phase feature from Eq.~\eqref{locnet_cnn} to get source-specific summary vectors. Here, source splitting is performed by masking instead of a direct mapping. Note that this masking is performed in a latent space and not in the frequency domain like done in cascaded systems like \cite{zqwang_localize, ziteng}. The summary vector will encode the DOA information specific to a source as the masks are used as weights to summarize information only from the corresponding source regions in the following way,

\begin{align}
    \label{summary_vector_wa}
     \xi^{n} (q) &= \frac{\sum_{t=1}^{T} w^{n} (t, q) z(t, q)}{\sum_{t=1}^{T} 
     w^{n} (t, q)},
\end{align}
where $\xi^{n} (q)$ is the summary vector for source $n$ at dimension $q$, $w^{n}(t, q) \in [0, 1]$ and $z(t, q) \in \mathbb{R}$ are the extracted feature mask (for source $n$) and the phase feature, respectively, at time $t$ and feature dimension $q$. 

The summary vector, represented in vector form as $\bm{\xi}^{n} \in \mathbb{R} ^{Q}$ is passed through a learnable source-specific $\text{AffineLayer}^{n}(\cdot)$, which acts as the predictor and converts the summary vector from dimension $Q$ to the dimension $\lfloor 360/\gamma \rfloor$, where $\gamma$ is the angle resolution in degrees to discretize the DOA angle. Based on this discretization, we can predict the DOA angle as a classifier with the softmax operation. From this, we can get the source-specific posterior probability for the possible angle classes as follows,
     
\begin{align}
     \label{posterior_2}
     [\Pr (\theta^{n}=\alpha_{i}|\mathcal{P})]_{i=1}^{\lfloor 360/\gamma \rfloor} &= \text{Softmax}(\text{AffineLayer}^{n}(\bm{\xi}^{n})), \\
     \label{class_angles_2}
     \alpha_{i} &=  \left( (\gamma * i) -\left((\gamma-1)/2\right) \right) \left(\pi/180\right),
\end{align}

The estimated DOA $\hat{\theta}^{n}$ for source $n$ is determined by finding the peak from the corresponding posterior in Eq.~\eqref{posterior_2} as follows,

\begin{align}
     \label{argmax_2}
     \hat{\theta}^{n} &= \operatorname*{argmax}_{\alpha_{i}} \Pr (\theta^{n}=\alpha_{i}|\mathcal{P}), 
\end{align}

In this architecture the predictor parameters are not shared across the sources in Eq.~\eqref{posterior_2} but it also possible to share them and an experimental comparison is discussed in Section \ref{sec:ps_exp}.

\subsection{Recurrent Mapping based Source Splitting Mechanism (Map-Split-R)}
\label{sec:blstm_ipd}
In this model, inter-microphone phase difference (IPD) \cite{mc_dc, bahmaninezhad2019comprehensive} is computed and passed as a feature directly to the model. There are no CNN layers in this model as pre-defined features are used. This model also performs a direct mapping like Section \ref{sec:cnn_blstm_map}. The IPD features are calculated as,

\begin{equation}
\label{ipd}
\overline{p}_{i}(t,f) = \frac{1}{M}  \lbrack \cos\angle (\frac{y_{i_1}(t,f)}{y_{i_2}(t, f)}) + \text{j} \sin\angle (\frac{y_{i_1}(t,f)}{y_{i_2}(t, f)}) \rbrack, i \in \{1, \dots, I\},
\end{equation}
where $y_m(t, f)$ is the input signal at channel $m$, time $t$ and frequency $f$,  $i$ represents an entry in a microphone pair list defined for calculating the IPD; and $i_1$ and $i_2$ are the index of microphones in each pair. We calculate IPD features for $I$ pairs and then concatenate their real and imaginary parts together. The concatenated IPD feature is represented as $\mathcal{\overline{P}} \in \mathbb{R} ^{T \times I \times 2F}$. The magnitude of the input signal at channel 1 i.e. $|Y_1|$ is also added as a feature to give the final input feature $\overline{Z} \in \mathbb{R} ^{T \times (2IF+F)}$. This feature is passed to a BLSTM component $\text{LocNet-BLSTM}(\cdot)$, which is the source splitter, as follows, 

\begin{align}
     [\overline{W}^{n}]_{n=1}^{N} &= \sigma(\text{LocNet-BLSTM} (\overline{Z})),
     \label{locnet_blstm}
\end{align}
where $\overline{W}^{n} \in \mathbb{R}^{T \times Q}$ is the source-specific feature for source $n$. This model achieves source splitting by a direct mapping so a simple average is used to extract the summary vector as follows,

\begin{align}
    \label{summary_vector}
     \overline{\xi}^{n} (q) &= \frac{1}{T} \sum_{t=1}^{T} \overline{w}^{n} (t, q),
\end{align}

The summary vector is used similar to Section \ref{sec:cnn_blstm} and the source specific DOAs are obtained by following Eq.~\eqref{posterior_2} and Eq.~\eqref{argmax_2}.

\subsection{Loss Functions}
\label{sec:loss_functions}
The multi-label classification model in Section \ref{sec:cnn_blstm_mlc} is trained with the binary cross entropy (BCE) loss. The models in Section \ref{sec:cnn_blstm} and Section \ref{sec:blstm_ipd} give source-specific probability distributions at the output. So cross entropy (CE) loss will be the natural choice. This choice will be good for a typical classification problem where the inter-class relationship is not important. As we are estimating the azimuth angles, we have classes that are ordered. So it is better to optimize by taking the class relationships into account as they are very informative. 

Typically a 1-hot vector is given as target to the CE loss. In DOA estimation it is better to predict a nearby angle compared to a far away angle. But the typical penalty from the CE loss doesn't take this into account. A loss function that takes the class ordering into account is the earth mover distance (EMD). EMD is the minimal cost required to transform one distribution to another \cite{levina_emd}. As in our problem the classes are naturally in a sorted order, we can use the simple closed-form solution from \cite{hou2016squared}. The solution is given by the squared error between the cumulative distribution function(CDF) of the prediction and the target probability distributions.

Another way to induce inter-class relationship is by using a soft target probability distribution which is not as sparse as the 1-hot vector. We use the following target distribution:

\begin{align}
\chi(i) = \begin{cases}
0.4 & i=\psi\\
0.2 & i=(\psi \pm 1) \mod \lfloor 360/\gamma \rfloor\\
0.1 & i=(\psi \pm 2) \mod \lfloor 360/\gamma \rfloor\\
0 & elsewhere
\end{cases}
\end{align}
where, $\chi(i)$ is the probability weight of the target distribution for class $i$ and $\psi$ is the index corresponding to the target angle class. When this soft target distribution is used with CE, we define it as the \textit{soft cross entropy} (SCE) loss. Similarly, it can also be used with EMD and we define it as  \textit{soft earth mover distance} (SEMD) loss. By assigning some probability weight to the angle classes in the neighbourhood of the target, the network can be potentially made to learn some inter-class relationship.

\section{Integration of Localization with ASR}
\label{sec:asr_integration}
MIMO-Speech \cite{chang2019mimo, mimo_transformer} is an end-to-end neural network that can perform simultaneous multi-talker speech recognition by taking the mixed speech multi-channel signal as the input. MIMO-Speech like typical ASR systems don't try to explicitly localize the sources. In this section we propose an extension in which MIMO-Speech uses source localization as a frontend. The azimuth angles of the all the sources estimated using the frontend is converted to a feature suitable for ASR and passed to MIMO-Speech as a localization prior. 

The MIMO-Speech network consists of three main components: (1) The masking subnetwork, (2) differentiable minimum variance distortionless response (MVDR) beamformer, and (3) the ASR subnetwork. The masking network in MIMO-Speech is monaural and gives channel-dependent time-frequency masks for beamforming. For this only the magnitude of the corresponding channel is given as input to the masking network. We will modify this masking network to take composite multi-channel input and output a channel-independent mask which will be shared across all channels. 

Firstly, it is possible to also get a multi-channel masking network baseline by using $\overline{Z}$ from Section \ref{sec:blstm_ipd} as the composite feature which is the concatenation of IPD features from Eq.~\eqref{ipd} and the magnitude of the first channel. For this to be further extended to take localization knowledge we need to concatenate it with an additional input feature that can encode the estimated azimuth angle. The groundtruth DOA is encoded as angle features in \cite{zhuo_loc, icassp-tencent,bahmaninezhad2019comprehensive}. We follow the same procedure but use the estimated DOA from our models proposed in Section \ref{sec:doa}. 

The steering vector $\mathbf{d}^{n} (f) \in \mathbb{C} ^{M}$ for source $n$ and frequency $f$ is calculated from estimated DOA $\hat{\theta}^{n}$. In this work we have used uniform circular arrays (UCA) and the steering vector is calculated as follows,

\begin{align}
     \tau_{m}^{n} &= \frac{r}{c} \cos({\hat{\theta}^{n} - \psi_{m}}),\quad m=1:M \\
     \mathbf{d}^{n} (f) &= [e^{j2\pi f \tau_{1}^{n}}, e^{j2\pi f \tau_{2}^{n}}, ..., e^{j2\pi f \tau_{M}^{n}}],
     \label{eq:svector}
\end{align}
where $\tau_{m}^{n}$ is the signed time delay between the $m$-th microphone and the center for source $n$, $\psi_{i}$ is the angular location of microphone $m$, $r$ is the radius of the UCA and $c$ is the speed of sound (343 m/s). 

The calculated steering vector is used to compute the angle features with the following steps,

\begin{align}
\tilde{a}^{n}(t,f) &= |\mathbf{d}^{n} (f)^\textrm{H} \mathbf{y} (t,f)|^{2}, \\
a^{n}(t,f) &= \tilde{a}^{n}(t,f) * \mathcal{I}(\tilde{a}^{n}(t,f) - \tilde{a}^{s}(t,f))_{s=1:N},
\label{eq:doa-mask-pre}
\end{align}
where $\mathbf{a}^{n} (t,f)$ is the angle feature for source $n$ at time $t$ and frequency $f$, $^\textrm{H}$ is conjugate transpose, $\mathbf{y} (t,f) \in \mathbb{C} ^{M}$ is the multichannel input, $N$ is the number of speakers, and $\mathcal{I}(.)$ is the indicator
function that outputs 0 if the input difference is negative for any of the ``$s=1:N$" cases and 1 otherwise. 

The computed angle features for all the sources are concatenated and included to the composite feature input list that is fed to the multi-channel masking network. This subnetwork given by $\text{MaskNet}(\cdot)$ gives the source specific masks $L^n \in (0,1)^{T \times F}$ as the output. Rest of the procedure is similar to the original MIMO-Speech model. The masks are used to compute the source specific spatial covariance matrix (SCM), $\mathbf{\Phi}^{n} (f)$  as follows:

\begin{align}
\mathbf{\Phi}^{n} (f) & = \frac{1}{\sum_{t=1}^T l^{n}(t,f)}\sum\limits_{t=1}^T l^{n}(t,f)\mathbf{y}(t, f)\mathbf{y}(t, f)^\textrm{H}, 
\label{psd}
\end{align}

The interference SCM, $\mathbf{\Phi}_{\mathsf{intf}}^{n} (f)$ for source $n$ is approximated as $\sum_{i\neq n}^N \mathbf{\Phi}^{i} (f)$ like \cite{chang2019mimo} (we experiment only with $N=2$, so no summation in that case). From the computed SCMs, the $M$-dimensional complex MVDR beamforming filter \cite{mvdr_souden} for source $n$ and frequency $f$, $\mathbf{b}^n (f) \in \mathbb{C} ^M$ is estimated as,

\begin{equation}
\mathbf{b}^n (f) = \frac{[\mathbf{\Phi}_{\mathsf{intf}}^{n} (f)+\mathbf{\Phi}_{\mathsf{noise}} (f)]^{-1} \mathbf{
\Phi}^{\text{n}} (f)}{\text{Tr}([\mathbf{\Phi}_{\mathsf{intf}}^{n} (f)+\mathbf{\Phi}_{\mathsf{noise}} (f)]^{-1}\mathbf{\Phi}^{\text{n}} (f))} \mathbf{u},
\label{estimate_mvdr_ref}
\end{equation}
where $\mathbf{u} \in \{0, 1\} ^{M}$ is a one-hot vector in which the index corresponding to the reference microphone is $1$, $\text{Tr}(\cdot)$ denotes the trace operation, and $\mathbf{\Phi}_{\mathsf{noise}} (f)$ is the noise SCM. The noise SCM can be estimated by obtaining an additional mask from the masking network. However we consider only stationary noise in this study and we experimentally found that it was better to ignore the noise SCM by considering it as an all-zero matrix.

With the estimated MVDR filter, we can perform speech separation to obtain the $n$-th separated STFT signal, $x^{n} (t, f) \in \mathbb{C}$ as follows:

\begin{equation}
 x^{n} (t, f) = \mathbf{b}^n (f)^\textrm{H} \mathbf{y}(t, f),
\label{apply_bf}
\end{equation}

This separated signal for source $n$, represented in matrix form as $X^{n} \in \mathbb{C}^{T \times F}$ is transformed to a feature suitable for speech recognition by performing log Mel filterbank transformation and utterance based mean-variance normalization (MVN). The extracted feature $O^{n}$ for source $n$ is passed to the speech recognition subnetwork $\text{ASR}(\cdot)$ to get $C^{n} = (c^{n}_{1}, c^{n}_{2}, \cdots)$, the token sequence corresponding to source $n$. The MIMO-Speech network is optimized with the reference text transcriptions $[C_{ref}^{i}]_{i=1}^{N}$ as the target. The joint connectionist temporal classification (CTC)/attention loss \cite{kim2017joint} is used as the ASR optimization criteria. 

Here again there is permutation ambiguity as there are multiple output sequences. We can solve it in two ways. First, we can follow the PIT scheme similar to the original MIMO-Speech model to resolve the prediction-target token sequence assignment problem. This takes additional computation time. When the PIT scheme is used, the order of the sources in the angle features in chosen randomly.  We propose to use the DOA knowledge to resolve the permutation ambiguity in the following way. During training stage, we can use the groundtruth DOA as input instead of the estimated DOA for computing the angle features. We determine the permutation of the target sequence based on the order of sources in which the angle features are concatenated and thereby can eliminate PIT. 

\section{Experiments}
\subsection{Data \& Setup}

\begin{table}[h]
  \centering
  \caption{The configurations used for simulating the 2-speaker mixtures}
    \begin{tabular}{l|l}
    \toprule
    \toprule
    \textit{\textbf{Simulation corpus}} & WSJCAM0 \\
    \midrule
    \textit{\textbf{ASR pretraining corpus}}  & WSJ \\
    \midrule
    \textit{\textbf{Sampling rate}} & 16 KHz \\
    \midrule
    \multirow{3}[0]{*}{\textit{\textbf{Num. of utterances}}} & Train - 7860  \\
          & Dev - 742 \\
          & Eval - 1088 \\
          \midrule
    \multirow{3}[0]{*}{\textit{\textbf{Num. of RIR}}}  & Train - 2620  \\
           & Dev - 742  \\
           & Eval - 1088 \\
    \midrule
    \textit{\textbf{T60}}   & $\mathcal{U}$(0.25s, 0.7s) \\
    \midrule
    \textit{\textbf{Num. of channels}} & 8 \\
    \midrule
    \textit{\textbf{Source distance from array}}  & $\mathcal{U}$(1m, 2m) \\
    \bottomrule
    \bottomrule
    \end{tabular}%
  \label{tab:simulation}%
\end{table}%

We simulated 2-speaker mixtures from WSJCAM0 \cite{Robinson1995WSJCAM0AB}. A uniform circular array (UCA) was used. Two types of mixtures with different array radius, 5cm (UCA-5) and 10cm (UCA-10) respectively, were simulated to study the impact of array aperture.  For UCA-10, background noise was added from REVERB corpus \cite{kinoshita2016summary} as it matches the array geometry used for recording that corpus and also to match the conditions of the real data used for evaluation. The signal-to-noise ratio (SNR) was uniformly sampled from the range $\mathcal{U}$(10dB, 20dB). The 5k vocabulary subset of the real overlapped data with stationary sources from MC-WSJ-AV was used just for evaluation with the models trained with the UCA-10 mixtures. Since the aperture cannot be changed on the real data we also experiment with choosing only the first three microphones of UCA-10. We call this a quadrant array (QA-10) in which the aperture is reduced as only the first quadrant from the circle is used. For each utterance, we mixed another utterance from a different speaker within the same set, so the resulting simulated data is the same size as the original clean data. The SMS-WSJ \cite{SmsWsj19} toolkit was used for creating the simulated data with maximum overlap. Image method \cite{allen1979image} was used to create the room impulse responses (RIR). Room configurations with the size (length-width-height) ranging from 5m-5m-2.6m to 11m-11m-3.4m were uniformly sampled while creating the mixtures. Both sources and the array are always chosen to be at the same height. The other configuration details used for simulation are shown in Table \ref{tab:simulation}. The source positions are randomly sampled and hence not constrained to be on a grid like \cite{chakrabarty2017multi}.

 A 25 ms window and a 10 ms frame shift with a Hanning window were used to compute STFT.  For the eight microphone configuration (UCA-10 and UCA-5) the following pair list: (1, 5), (2, 6), (3, 7), (4, 8), (1, 3), (3,5), (5,7), and (7, 1) were used to compute the IPD features defined in Eq~\eqref{ipd}. The pair list for the three microphone subset (QA-10) was: (1, 2), (2, 3), and (1, 3). Three CNN blocks with rectified linear unit (ReLU) activation after each block followed by a feedforward layer were used as $\text{LocNet-CNN}(\cdot)$ defined in Eq.~\eqref{locnet_cnn}. The CNN filters are applied across the microphone-frequency dimensions with a stride of $1$. The kernel shapes were dependent on the number of input microphones. In the case of 8-mics (UCA-10 and UCA-5), kernels of shapes, $4 \times 1$, $3 \times 3$, and $3 \times 3$ were used for the first, second, and third CNN block respectively. For the 3-mic QA-10, kernels of shapes, $2 \times 1$, $2 \times 3$, and $1 \times 3$ were used for the first, second, and third CNN block respectively. Padding was not performed for the channel dimension so as to fuse them after the final CNN block. The number of feature maps were
4 in the first block, 16 in the second block, and 32 in the final block. $Q$ was fixed as ``$2 \times \lfloor 360/\gamma \rfloor$".  One output gate projected
bidirectional long short-term memory (BLSTMP) layer with $Q$ cells was used as $\text{LocNet-Mask}(\cdot)$ defined in Eq.~\eqref{locnet_mask}. All the DNN based multi-source localization were optimized  with Adam \cite{kingma2015adam} for 50 epochs and a learning rate of $10^{-3}$.

The masking network for MIMO-Speech was designed with two BLSTMP layers with $771$ cells. The encoder-decoder ASR network was based on the Transformer architecture \cite{espnet_transformer} and it was initialized with a pretrained model that used single speaker training utterances from both WSJ0 and WSJ1. The same architecture as \cite{mimo_transformer} was used for the encoder-decoder ASR
model. It has 12 layers in the encoder and 6 layers in the
decoder. Before the Transformer encoder, the log Mel filterbank features of $80$ dimensions are encoded by two CNN blocks. The CNN layers have a kernel size of $3 \times 3$ and the number of feature maps is
64 in the first block and 128 in the second block.  Attention/CTC joint ASR decoding was performed with score combination with a word-level recurrent language model from \cite{wordlm} trained on the text data from WSJ. Our implementation was based on the Pytorch backend of ESPnet \cite{espnet}.

 The input signal was preprocessed with weighted prediction error (WPE) \cite{wpe, Drude2018NaraWPE} based dereverberation with a filter order of $10$ and prediction delay $3$, only during inference time . The second channel was fixed as the reference microphone for MVDR in Eq.~\eqref{estimate_mvdr_ref}. We evaluate several localization models with the proposed DOA integration with MIMO-Speech. To avoid retraining every time, the groundtruth DOA was used while training the network and DOA estimation was performed only while inference. While PIT training, the order of the angle features was randomly permuted. Subspace DOA estimation was performed with ``Pyroomacoustics" toolkit \cite{pra}.

\subsection{Multi-Source Localization Performance}

\begin{table}[H]
  \centering
  \caption{Mean absolute error (MAE) in degrees on both our UCA-5 and UCA-10 simulated 2-source mixtures comparing our proposed source-splitting methods with subspace methods, and multi-label classification. The symbol ``\xmark" \ implies that the option is not used and ``\cmark" \ implies that it is used.}
   \begin{adjustbox}{width=\columnwidth,center}
    \begin{tabular}{c|c|c|c|c|c|c|c|c|c|c|c|c}
    \toprule
    \toprule
    \multirow{2}[1]{*}{\textit{\textbf{Row}}} &
    \multirow{3}[2]{*}{\textit{\textbf{Method}}} & 
    \multirow{3}[2]{*}{\textit{\textbf{$\gamma$}}} &
    \multirow{2}[1]{*}{\textit{\textbf{Loss}}} &\multirow{3}[2]{*}{\textit{\textbf{PIT}}} & \multicolumn{4}{c|}{\textit{\textbf{UCA-5 (M=8)}}} &\multicolumn{4}{c}{\textit{\textbf{UCA-10 (M=8)}}} \\
\multirow{2}[1]{*}{\textit{\textbf{ID}}}  &  &     & \multirow{2}[1]{*}{\textit{\textbf{Function}}} & & \multicolumn{2}{c|}{\textit{\textbf{No WPE}}} & \multicolumn{2}{c|}{\textit{\textbf{WPE}}} & \multicolumn{2}{c|}{\textit{\textbf{No WPE}}} & \multicolumn{2}{c}{\textit{\textbf{WPE}}} \\
&   &       &  & &\textit{\textbf{Dev}} & \textit{\textbf{Test}} & \textit{\textbf{Dev}} & \textit{\textbf{Test}} & \textit{\textbf{Dev}} & \textit{\textbf{Test}} & \textit{\textbf{Dev}} & \textit{\textbf{Test}} \\
    \midrule
    \midrule
    1 & MUSIC &   1 & - & - & 32.9 & 28.3 & 21.0 & 16.8 & 17.0 & 14.1 & 11.0 & 8.7\\
   2 &  MUSIC-NAM &  1 & - &  - & 8.7  & 6.8 & 6.8 & 4.8 & 2.5 & 2.2 & 2.1 & 1.1\\

    3 & TOPS &  1 & - & - & 7.9  & 6.9 & 7.4  &  6.3 & 3.1 & 2.6 & 1.9 & 1.8 \\
    4 & MLC &  5 & BCE  & - & 18.1 & 18.2 & 22.0 & 20.0 & 19.9 & 21.6 & 20.8 & 20.7\\
    5 & MLC &  10 & BCE  & - & 10.5 & 10.9 & 13.1 & 12.7 & 7.8  & 8.4 & 9.1 & 9.0\\
    \midrule
    6 & Map-Split-C & 10 & CE & \xmark & 4.6 & 4.8 & 4.0 & 3.9 & 5.3 & 5.8 & 5.3 & 6.2\\
    7 & Mask-Split &  10 & CE  & \xmark & 2.9 & 3.3 & 3.1 & 3.2  & 4.0 & 4.7 & 3.3 & 3.4\\
    8 & Mask-Split &  1 & CE  & \xmark & 28.7 & 27.0 & 21.0 & 19.2 & 32.9 & 34.1 & 28.9 & 27.4\\
    9 & Mask-Split &  1 & SCE  & \xmark & 1.6 & 1.9 & 2.4 & 2.0  & 3.7 & 3.6 & 3.6 & 3.8 \\
    10 & Mask-Split &  1 & EMD  & \xmark & 4.7 & 4.4 & 4.8 & 4.7  & 3.2 & 3.3 & 3.6 & 3.6 \\
    11 & Mask-Split &  1 & SEMD  & \cmark & 3.0 & 3.0 & 4.0 & 3.9  & 2.8 & 2.6 & 2.8 & 2.9 \\
    12 & Mask-Split &  1 & SEMD  & \xmark & \textbf{1.4} & \textbf{1.5} & \textbf{1.4} & \textbf{1.4}  & \textbf{1.5} & \textbf{1.4} & \textbf{1.1} & \textbf{1.1} \\
    13 & Map-Split-C &  1 & SEMD  & \cmark & 3.7 & 4.2 & 3.8 & 4.1 & 3.6 & 4.0 & 3.2 & 3.8\\
    14 & Map-Split-C &  1 & SEMD  & \xmark & 3.8 & 4.5 & 3.6 & 4.3 & 4.0 & 4.1 & 3.6 & 3.9\\
    15 & Map-Split-R &  1 & SEMD  & \cmark & 2.7 & 2.9 & 2.8 & 3.1    & 2.3 & 2.2 & 2.2 & 2.5 \\
    16 & Map-Split-R &  1 & SEMD  & \xmark & 2.0 & 2.1 & 2.2 & 2.2  & 2.1 & 2.0 & 2.3 & 2.1\\
    \bottomrule
    \bottomrule
    \end{tabular}%
     \end{adjustbox}
  \label{tab:doa_error}%
  
\end{table}%

\begin{table}[H]
  \centering
  \caption{Results on the simulated quadrant array (QA-10) 2-source mixtures comparing our proposed source-splitting methods with subspace methods, and multi-label classification. QA-10 corresponds to using only the first three microphones of the eight microphone UCA-10. MAE in degrees is the metric. The signals were preprocessed with WPE to remove late reverberations. The symbol ``\xmark" \ implies that the option is not used and ``\cmark" \ implies that it is used.}
    \begin{tabular}{c|c|c|c|c|c|c}
    \toprule
    \toprule
    \multirow{2}[1]{*}{\textit{\textbf{Row ID}}} &
    \multirow{2}[1]{*}{\textit{\textbf{Method}}} & 
    \multirow{2}[1]{*}{\textit{\textbf{$\gamma$}}} &
    \multirow{2}[1]{*}{\textit{\textbf{Loss Function}}} &\multirow{2}[1]{*}{\textit{\textbf{PIT}}} & \multicolumn{2}{c}{\textit{\textbf{QA-10 (M=3)}}}  \\
  &  &     &  & & \textit{\textbf{Dev}} & \textit{\textbf{Test}} \\
    \midrule
    \midrule
    1 & MUSIC &   1 & - & - & 58.7 & 57.7 \\
   2 &  MUSIC-NAM &  1 & - &  - & 22.7  & 21.7\\

    3 & TOPS &  1 & - & - & 10.6 & 9.7 \\
    \midrule
    4 & MLC &  10 & BCE  & - & 10.1 & 11.9 \\
    \midrule
    5 & Mask-Split &  1 & SEMD  & \xmark & \textbf{2.5} & \textbf{2.8} \\
    6 & Map-Split-C &  1 & SEMD  & \xmark & 4.0 & 4.4\\
    7 & Map-Split-R &  1 & SEMD  & \xmark & 2.8 & 3.5\\
    \bottomrule
    \bottomrule
    \end{tabular}%
  \label{tab:doa_error_qa}%
  
\end{table}%

We use three popular subspace-based signal processing methods MUSIC, MUSIC with normalized arithmetic mean fusion (MUSIC-NAM) \cite{music_nam}, and TOPS as baselines. Frequencies from 100 Hz to 8000 Hz were used for the subspace methods. For all three, the spatial response is computed and the top two peaks are detected to estimate the DOA. The average absolute cyclic angle difference between the predicted angle and the ground-truth angle in degrees is used as the metric. The permutation of the prediction with the reference that gives the minimum error is chosen. 

The results with and without WPE preprocessing for both UCA-5 and UCA-10 mixtures  are given in Table \ref{tab:doa_error}.  Most of the supervised deep learning models with different configurations perform significantly better than the subspace methods for UCA-5 but for UCA-10 both MUSIC-NAM and TOPS give very strong results. The results show that the deep learning methods are robust to reverberations and give good results even without the WPE frontend.

The results of the multi-label classification model from Section \ref{sec:cnn_blstm_mlc} are given in rows 4 \& 5. We can see that a higher resolution of $\gamma=5$ degrades the performance from $\gamma=10$. This is because of spurious peaks in nearby angles when the resolution is increased. This shows the need for having separate output vectors for different sources. Rows 6-16 gives the results of the models with the proposed source splitting mechanism. The results  are given with both PIT and also fixing the targets in ascending order. From the results we can see that fixing the target order works better. 

Rows 8-12 use the Mask-Split model proposed in Section \ref{sec:cnn_blstm} with phase feature masking. The CE loss with  $\gamma=10$ (row 7) works reasonably well and gives an improvement over the multi-label classification model. Increasing the resolution to $\gamma=1$  (rows 8) makes it poor because of its inability to learn inter-class relationship like explained in Section \ref{sec:loss_functions}. Making the target distribution smoother with SCE loss alleviates the problem quite well and we can observe better results in row 9. Using EMD loss also has a similar effect (row 10). Combining both ideas with the SEMD loss gives the best performance with a prediction error of less than or equal to $1.5^{\circ}$ when PIT is not used (row 12) for both array configurations. The results of both Map-Split models (row 13-row16) are a bit worse compared to the Mask-Split model and amongst them Map-Split-R works better.

The results on the quadrant array with WPE preprocessing are shown in Table \ref{tab:doa_error_qa}. This again shows that subspace methods degrade severely when the aperture is reduced. Here the proposed methods significantly outperform both subspace methods and MLC.

\subsubsection{Predictor Sharing}
\label{sec:ps_exp}
\begin{table}[ht]
  \centering
  \caption{MAE (degree) on the simulated UCA-10 2-source mixtures demonstrating the effect of sharing the predictor parameters across both the sources on the three different proposed source-splitting methods. The signals were preprocessed with WPE to remove late reverberations. The symbol ``\xmark" \ implies that the option is not used and ``\cmark" \ implies that it is used.}
    \begin{tabular}{c|c|c|c|c}
    \toprule
    \toprule
    \multirow{2}[1]{*}{\textit{\textbf{Method}}} & 
    \multirow{2}[1]{*}{\textit{\textbf{Predictor Sharing}}} &\multirow{2}[1]{*}{\textit{\textbf{PIT}}} & \multicolumn{2}{c}{\textit{\textbf{UCA-10}}}  \\
  &  &     & \textit{\textbf{Dev}} & \textit{\textbf{Test}} \\
    \midrule
    \midrule
    Map-Split-C  & \cmark & \cmark & 3.2 & 3.8 \\
    Map-Split-C  & \cmark & \xmark & 3.6 & 3.9\\
\cdashline{1-5}
    Map-Split-C  & \xmark & \cmark & 5.2 & 5.5\\
    Map-Split-C  & \xmark & \xmark & 3.5 & 4.1\\
    \midrule
    Map-Split-R & \cmark & \cmark & 5.7 & 5.4\\
    Map-Split-R & \cmark & \xmark & 5.7 & 5.2\\
\cdashline{1-5}
    Map-Split-R & \xmark & \cmark & 2.2 & 2.5 \\
    Map-Split-R & \xmark & \xmark &  2.3 & 2.1 \\
    \midrule
    Mask-Split & \cmark & \cmark & 6.0 & 6.0 \\
    Mask-Split & \cmark & \xmark & 2.3 & 2.2 \\
\cdashline{1-5}
    Mask-Split & \xmark & \cmark & 2.8 & 2.9 \\
    Mask-Split & \xmark & \xmark & 1.1 & 1.1 \\
    \bottomrule
    \bottomrule
    \end{tabular}%
  \label{tab:doa_error_ps}%
  
\end{table}%

In Table \ref{tab:doa_error_ps}, a comparison is made between sharing the parameters of the predictor (source independent) and a source dependent predictor for all the three proposed source splitting methods. The results are shown for UCA-10 but a similar trend was observed for the other array configurations too. We can see that having a source dependent predictor benefits Mask-Split and Map-Split-R but not Map-Split-C. So by default Mask-Split and Map-Split-R make the predictor feedforward layer dependent on source $n$ as given in Eq~\eqref{posterior_2} but Map-Split-C makes this layer independent of the source as given in Eq~\eqref{posterior_1}.

\subsubsection{Impact of SNR}

\begin{table}[ht]
  \centering
  \caption{Results on the simulated UCA-10 2-source mixtures demonstrating the effect of SNR. MAE in degrees is the metric. The signals were preprocessed with WPE to remove late reverberations. PIT was not used for the source splitting methods in these experiments.}
    \begin{tabular}{c|c|c|c|c|c|c}
    \toprule
    \toprule
    \multirow{2}[1]{*}{\textit{\textbf{Row}}} &
    \multirow{2}[1]{*}{\textit{\textbf{Method}}} & 
    \multirow{2}[1]{*}{\textit{\textbf{$\gamma$}}} &
    \multirow{2}[1]{*}{\textit{\textbf{Loss}}} &\multirow{2}[1]{*}{\textit{\textbf{PIT}}} & \multicolumn{2}{c}{\textit{\textbf{UCA-10}}}  \\
  \multirow{2}[1]{*}{\textit{\textbf{ID}}} &  &     & \multirow{2}[1]{*}{\textit{\textbf{Function}}} & & \textit{\textbf{SNR - 0dB}} & \textit{\textbf{SNR - 30dB}} \\
 &&&& & \textit{\textbf{Test}} & \textit{\textbf{Test}} \\
 \midrule
    \midrule
    1 & MUSIC &   1 & - & - & 14.2 & 15.6 \\
   2 &  MUSIC-NAM &  1 & - &   - & 1.3 & 1.1 \\

    3 & TOPS &  1 & - & - & 2.3 & 1.6 \\
    \midrule
    4 & MLC &  10 & BCE  & - & 8.9 & 10.3 \\
    \midrule
    5 & Mask-Split &  1 & SEMD  & \xmark & 1.5 & 1.8 \\
    6 & Map-Split-C &  1 & SEMD  & \xmark & 6.0 & 4.0\\
    7 & Map-Split-R &  1 & SEMD  & \xmark & 2.8 & 2.2 \\
    \bottomrule
    \bottomrule
    \end{tabular}%
  \label{tab:doa_error_snr}%
  
\end{table}%

Two additional test sets were simulated for the UCA-10 configuration by fixing the SNR as 0 dB and  30 dB respectively to see the impact of SNR on DOA estimation.  We can see from Table \ref{tab:doa_error_snr} that SNR doesn't play a significant role in the DOA estimation performance for most of the DOA estimation methods. Only, Map-Split-C seems to have a notable degradation for the challenging 0 db SNR scenario. The background noise that we used from REVERB \cite{kinoshita2016summary} consists mostly of stationary noise so both the baselines and the proposed methods are quite robust to it. 

\subsubsection{Impact of Angular Distance between Sources on DOA Estimation}
\begin{table}[ht]
  \centering
  \caption{Results on the simulated UCA-5 2-source mixtures demonstrating the impact of angular distance between the two sources. The results here are divided into four subsets based on the angle difference. MAE in degrees is the metric. The signals were preprocessed with WPE to remove late reverberations. PIT was not used for the source splitting methods in these experiments.}
   \begin{adjustbox}{width=\columnwidth,center}
    \begin{tabular}{c|c|c|c|c|c|c|c|c|c|c|c|c}
    \toprule
    \toprule
    \multirow{2}[1]{*}{\textit{\textbf{Row}}} &
    \multirow{2}[1]{*}{\textit{\textbf{Method}}} & 
    \multirow{2}[1]{*}{\textit{\textbf{$\gamma$}}} &
    \multirow{2}[1]{*}{\textit{\textbf{Loss}}} &\multirow{2}[1]{*}{\textit{\textbf{PIT}}} & \multicolumn{8}{c}{\textit{\textbf{UCA-5}}}  \\
  \multirow{2}[1]{*}{\textit{\textbf{ID}}} &  &     & \multirow{2}[1]{*}{\textit{\textbf{Function}}} & & \multicolumn{4}{c|}{\textit{\textbf{Dev}}} & \multicolumn{4}{c}{\textit{\textbf{Test}}} \\
 &&&&& \textit{\textbf{10-20}} & \textit{\textbf{21-45}} & \textit{\textbf{46-90}} & \textit{\textbf{91-180}}& \textit{\textbf{10-20}} & \textit{\textbf{21-45}} & \textit{\textbf{46-90}} & \textit{\textbf{91-180}} \\
 \midrule
    \midrule
    1 & MUSIC &   1 & - & - & 42.8 & 23.9 & 13.9 & 6.7 & 36.0 & 16.8 & 14.2 & 7.0 \\
   2 &  MUSIC-NAM &  1 & - &   - & 67.0 & 8.0 & \textbf{0.4} & \textbf{0.5} & 65.3 & 4.7 & \textbf{0.5} & \textbf{0.5} \\

    3 & TOPS &  1 & - & - & 67.6 & 8.3 & 1.3 & 1.1 & 67.3 & 9.2 & 1.6 & 1.1\\
    \midrule
    4 & MLC &  10 & BCE  & - & 25.6 & 12.9 & 11.0 & 12.5 & 19.9 & 15.4 & 9.2 & 12.7 \\
    \midrule
    5 & Mask-Split &  1 & SEMD  & \xmark & \textbf{1.6} & \textbf{1.4} & 1.3 & 1.4 & \textbf{1.9} & \textbf{1.4} & 1.2 & 1.5\\
    6 & Map-Split-C &  1 & SEMD  & \xmark & 5.5 & 4.6 & 3.3 & 3.2 & 5.8 & 5.0 & 4.7 & 3.8\\
    7 & Map-Split-R &  1 & SEMD  & \xmark & 3.6 & 2.5 & 1.9 & 2.1 & 3.4 & 2.9 & 2.2 & 1.9\\
    \bottomrule
    \bottomrule
    \end{tabular}%
     \end{adjustbox}
  \label{tab:doa_error_ad}%
  
\end{table}%

In \cite{icassp-tencent, bahmaninezhad2019comprehensive}, it was shown that it is very challenging to separate sources when the angular distance between the sources is very small. The results in Table \ref{tab:doa_error} for the UCA-5 array showed that the proposed source splitting method significantly outperforms the subspace methods. In Table \ref{tab:doa_error_ad}, we further investigate the results on the UCA-5 configuration by checking the performance based on the angular distance between the two sources (speakers). Both the development and test sets were divided into four subsets based on the angle difference in degrees: (1) ``10-20", (2) ``21-45", (3) ``46-90", and (4) ``91-180". We can see the performance of all three subspace methods (rows 1-3) are very poor in the most challenging scenario of ``10-20" and they also have a significant degradation in the ``21-45" cases. The MLC baseline (row 4) also degrages when the angle difference is  ``10-20" but it is relatively better than the subspace methods. The proposed source splitting methods (rows 5-7) are very robust to the angular distance between the sources and amongst them Mask-Split is the most robust. 

\subsection{ASR Performance}

\begin{table}[htbp]
  \centering
  \caption{ASR performance on the simulated 2-speaker mixtures comparing our proposed DOA integration method with vanilla MIMO-Speech. The input signal was preprocessed with WPE to remove late reverberations. Word error rate (WER) is used as the metric for comparison. For WER, lower the better. The symbol ``\xmark" \ implies that the option is not used and ``\cmark" \ implies that it is used.}

   \resizebox{\linewidth}{!}{%
    \begin{tabular}{c|c|c|c|c|c|c|c|c|c|c|c}
    \toprule
    \toprule
          &\textit{\textbf{Row}} & \textit{\textbf{DOA}} & \textit{\textbf{DOA}} & \textit{\textbf{IPD}} &
          \textit{\textbf{ASR}} &\multicolumn{2}{c|}{\textit{\textbf{UCA-5}}} &\multicolumn{2}{c|}{\textit{\textbf{UCA-10}}} &\multicolumn{2}{c}{\textit{\textbf{QA-10}}} \\ 
          
          & \textit{\textbf{ID}} & \textit{\textbf{Method}}
          & \textit{\textbf{PIT}} & \textit{\textbf{for ASR}}
          & \textit{\textbf{PIT}} & \textit{\textbf{Dev}} & \textit{\textbf{Test}} & \textit{\textbf{Dev}} & \textit{\textbf{Test}} & \textit{\textbf{Dev}} & \textit{\textbf{Test}} \\
    \midrule
        \midrule
    Clean  & 1   &- &-  & \xmark  & \xmark   &  13.7 & 13.6  & 13.7 & 13.6 & 13.7 & 13.6 \\
    Input mixture (ch-1) & 2  &- &- & \xmark  & \xmark  & 111.6 & 115.6 & 112.1 & 116.3 & 112.1 & 116.3\\
    \midrule
    \textit{Oracle} binary Mask (IBM) & 3  &- &- & \xmark   & \xmark     & 6.6 & 6.1  & 7.1 & 6.5 & 12.7 & 11.8\\
    \midrule
     \multirow{2}[2]{*}{Baseline MIMO-Speech} & 4 & - & - & \xmark  & \cmark  &  14.0  &  12.0 & 15.1 & 13.6 & 24.8 & 22.2 \\
     & 5  & - & - & \cmark   & \cmark   & 10.4 & 9.2 & 13.1 & 11.5 & 24.0 & 22.0 \\
    \midrule
    \multirow{2}[2]{*}{MIMO-Speech + \textit{Oracle} Angle Feature}  & 6  & - & -  & \cmark   & \cmark      & 6.0 & 5.7 & 6.8 & 6.0 & 11.9 & 10.7 \\
     & 7  & - & - &\cmark  &  \xmark    & 6.3 & 5.6 & 6.6 & 6.2 & 12.0 & 11.1\\
    \midrule
    \multirow{10}[5]{*}{MIMO-Speech + DOA Estimation}  & 8  & MUSIC & -  & \cmark & \xmark      & 16.9 & 14.2 & 12.9 & 11.5 & 27.2 & 25.2  \\
     & 9  & MUSIC-NAM & -  & \cmark & \xmark      & 12.0 & 9.7 & 8.6 & 7.4 & 25.5 & 24.1 \\
     & 10  & TOPS & - & \cmark  & \xmark      & 12.6 & 11.3 & 8.0 & 7.2 & 17.3 & 15.7  \\
    & 11  & MLC & - & \cmark  & \xmark      & 13.1 & 11.3 & 11.6 & 11.5 & 15.5 & 15.6  \\
    & 12  & Map-Split-C & \xmark & \cmark  & \xmark      & 7.8 & 7.7 & 9.1 & 8.6 & 14.6 & 13.3  \\
     & 13  & Map-Split-R & \cmark & \cmark  & \cmark      & 6.9 & 6.2 & 7.3 & 6.8 & 12.4 & 11.5  \\
     & 14 & Map-Split-R & \xmark & \cmark   & \cmark      & 6.6 & 6.2 &7.3 & 6.7  & 12.6 & 11.8 \\
     & 15 & Map-Split-R & \xmark & \cmark   & \xmark      & 6.8 & 6.1 & 7.2 & 6.9  & 12.7 & 12.1 \\
     & 16  & Mask-Split & \xmark & \cmark  & \cmark      & 6.1 & 5.9 & 7.0 & 6.3 & 12.5 & 11.7  \\
     & 17  & Mask-Split & \xmark & \cmark  &  \xmark    & 6.3 & 6.0 &7.0 & 6.3 & 13.0 & 11.8 \\
    \bottomrule
    \bottomrule
    \end{tabular}%
    }
  \label{tab:wer}%
\end{table}%

The speech recognition performance of our proposed DOA integration compared with the vanilla MIMO-Speech baseline is given in Table \ref{tab:wer}. Word error rate (WER) is used as the metric for ASR and the results are shown for mixtures from all three array configurations.  The results of the clean single-speaker data with the pretrained ASR model is given in row 1. Note that the results of the clean single-speaker data  is not very good because it is from the British English corpus WSJCAM0 and the pretrained model was trained with American English data from WSJ. This accent difference is fixed in the mixed speech models as the ASR network will be fine tuned in that case with British English data. The ASR results of the simulated mixture using the single-speaker model is also shown (WER more than 100\% because of too many insertion errors).  

Oracle experiment by directly giving the reference ideal binary masks (IBM) to the beamformer is given in row 3. Row 4 shows the results of the baseline MIMO-Speech with the architecture originally proposed in \cite{mimo_transformer}. Row 5 shows the modified baseline by adding IPD features and a channel independent masking network as mentioned in Section \ref{sec:asr_integration}. This makes the baseline stronger and gives slightly better results. Oracle ASR-DOA integration experiments were performed by using the groundtruth DOA while inference as a proof of concept and the results are shown in rows 6 \& 7. We can see a very significant improvement in this case with the word error rates  reduced by close to a factor of two. The results are also slightly better than using the oracle binary masks. This proves the importance of feeding localization knowledge to a multi-speaker ASR system. We can also see that turning off PIT and fixing the target sequence order based on the DOA order gives similar performance (row 7) with added benefits. This not only saves computation time while training but also makes the inference more informative by associating the DOA and its corresponding transcription.

In rows 8-17 some of the DOA estimation methods are used to compute the angle features. All the methods except MLC here are high resolution and use $\gamma=1$. For MLC $\gamma=10$ was used. We can see that using MUSIC, MUSIC-NAM, TOPS, or MLC degrades the ASR performance from the baseline for the lower aperture scenarios of UCA-5 and QA-10. This shows that localization knowledge will be useful only if we can estimate with good precision and reliability. The results with the proposed deep learning based DOA estimation and SEMD loss are shown in rows 12-17. With any of these methods we get results close to using the oracle DOA. We can see the Mask-Split model which gives the best localization performance in Table \ref{tab:doa_error} also gives the best results here (row 14). This result almost matches the performance obtained with oracle masks in row 3.

\subsubsection{Real Data}

\begin{figure}[H]
\centering
\includegraphics[scale=0.5]{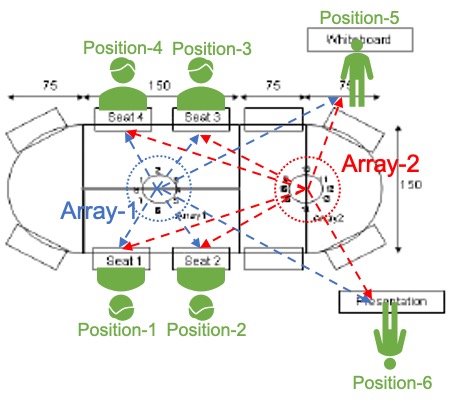}
\caption{Layout of MC-WSJ-AV}
\label{fig:layout}
\end{figure}

The overlapped data from MC-WSJ-AV corpus \cite{mc_wsj} was also used for ASR evaluation. The positions of the speakers are mentioned to be stationary throughout the utterance. A schematic diagram of the recording layout redrawn from \cite{mc_wsj} is shown in Figure \ref{fig:layout}. There are six possible positions in the room that the speakers were positioned. As there are two speakers in the mixture, there are fifteen possible position pairs. The real data has recordings from two arrays with the same configuration placed in different positions. From the schematic diagram we can see that four of the six positions are very close to array-1 and all positions are generally far from Array-2. The model trained with the simulated UCA-10 and QA-10 mixtures are used here as they follow similar configurations. Although trained with simulated data, it is crucial to see how our methods work on real data. As there are no groundtruth DOA labels, we cannot evaluate the angle prediction error.

\begin{table}[H]
  \centering
  \caption{ASR performance on the real 2-speaker overlapped data from MC-WSJ-AV. The input signal was preprocessed with WPE to remove late reverberations. Word error rate (WER) is used as the metric for comparison. For WER, lower the better.}
       \begin{adjustbox}{width=\columnwidth,center}
    \begin{tabular}{c|c|c|c|c|c}
    \toprule
    \toprule
     &  & \multicolumn{2}{c|}{\textit{\textbf{UCA-10 (M=8)}}}& \multicolumn{2}{c}{\textit{\textbf{QA-10 (M=3)}}} \\
          & \textit{\textbf{DOA Method}} & \textit{\textbf{Array-1}} & \textit{\textbf{Array-2}} & \textit{\textbf{Array-1}} & \textit{\textbf{Array-2}} \\
    \midrule
    \midrule
    MIMO-Speech & N/A   & 34.3  & 47.6  & \textbf{41.6} &  \textbf{55.7} \\
    MIMO-Speech w/ IPD & N/A   & \textbf{20.6}  & \textbf{30.7}  & 48.1 & 63.1 \\
    \midrule
    MIMO-Speech + DOA Integration & MUSIC-NAM &  10.7    &  \textit{\textbf{17.1}}     &  36.4 & 46.8 \\
    MIMO-Speech + DOA Integration & TOPS  &   \textit{\textbf{10.6}}   &   19.2    & \textbf{20.0} & \textbf{33.4} \\
    \midrule
    MIMO-Speech + DOA Integration & Mask-Split & 14.0  & 18.2  & \textit{\textbf{18.0}} & 28.7 \\
MIMO-Speech + DOA Integration & Map-Split-R &  14.8 & 22.3 & 21.2 & 25.6 \\
    MIMO-Speech + DOA Integration & Map-Split-R (chunking) & \textbf{12.1}  & \textbf{17.3}  & 20.7 & \textit{\textbf{25.3}} \\
    \bottomrule
    \bottomrule
    \end{tabular}
    \end{adjustbox}
  \label{tab:wer_real}%
\end{table}%

The ASR performance is shown in Table \ref{tab:wer_real}. The DOA estimation here is performed with $\gamma=1$ and the source splitting models here are without PIT and with SEMD loss. We also experiment with a chunking scheme here for Map-Split-R. Here we splice the input audio into 100 ms chunks with 50\% overlap. DOA estimation is performed on each of these chunks and their median output is chosen as the estimated DOA. This was done to remove the effect of some bad chunks in the estimation process. 

Using estimated DOAs from either the subspace methods or the source splitting models, ASR-DOA integration significantly outperforms the MIMO-Speech baselines for this data. This further proves the importance of localization as a very pivotal frontend for ASR. As positions of the sources in this dataset were chosen to be favorable to array-1 it gives better results. For UCA-10, there is not much difference between the subspace methods and the proposed source splitting methods but for QA-10, the proposed method outperforms and the margin is significant for array-2. This again shows the effectiveness of the source splitting approach on challenging conditions and lower aperture. 

\section{Conclusion \& Future Work}
\vspace*{-2mm}
\label{sec:conclusion}
We proposed a novel deep learning based model for multi-source localization that can classify DOAs with a high resolution. An extensive evaluation was performed with different choices of architectures, loss functions, classification resolution, and training schemes that handle permutation ambiguity. Our proposed source splitting model was shown to have a significantly lower prediction error compared to a multi-label classification model. The proposed method also outperforms well known subspace methods. We also proposed a soft earth mover distance (SEMD) loss function for the localization task that models inter-class relationship well for DOA estimation and hence predicts near perfect DOA. 

We have also devised a method to use the proposed DOA estimation as a frontend for multi-talker ASR. This integration greatly helps speech recognition and shows the importance of using localization priors with far-field ASR models. Based on this finding, ASR performance was used as the metric of evaluation for DOA estimation in real data where DOA labels are not available to compute prediction errors.

In future, we would like to extend our methods to also handle source counting within the model to make it work for arbitrary number of sources. One possible approach would be to use a conditional chain model that is popular for source separation \cite{rsh, shi2020sequence}. The other important extension is adapting our method to work on more challenging and realistic data like CHiME-6 \cite{watanabe2020chime6} which involves a complicated distributed array setup with moving sources.  

\bibliography{mybibfile}

\end{document}